\documentclass[10pt,twocolumn]{IEEEtran}
\usepackage{graphicx}           
\usepackage{amsmath,amsfonts}
\usepackage{times,epsfig}
\usepackage{psfrag}
\usepackage{subfigure}
\usepackage{stfloats}
\usepackage{amssymb}
\usepackage{multirow}
\usepackage[justification=centering]{caption}
\usepackage{cite}
\usepackage{CJK}
\usepackage{url}
\usepackage{caption}
\usepackage{mathrsfs}
\usepackage{bm}
\usepackage{array}
\usepackage{algorithmic}
\usepackage[lined,ruled,linesnumbered]{algorithm2e}

\begin{document}
\pagenumbering{arabic}
\bibliographystyle{ieeetr}

\title{Joint User Association and Resource Pricing for Metaverse: Distributed and Centralized Approaches}
\author{
	\IEEEauthorblockN{Xumin Huang\IEEEauthorrefmark{1}, Weifeng Zhong\IEEEauthorrefmark{1}, Jiangtian Nie\IEEEauthorrefmark{2}, Qin Hu\IEEEauthorrefmark{3}, Zehui Xiong\IEEEauthorrefmark{4}, \\Jiawen Kang\IEEEauthorrefmark{1}, and Tony Q. S. Quek\IEEEauthorrefmark{4}}\\
	\IEEEauthorblockA{\IEEEauthorrefmark{1}School of Automation, Guangdong University of Technology, Guangzhou, China\\
		\IEEEauthorrefmark{2}School of Computer Science and Engineering, Nanyang
		Technological University, Singapore\\
		\IEEEauthorrefmark{3}Department of Computer and Information Science, Indiana
		University-Purdue University Indianapolis, USA\\
		\IEEEauthorrefmark{4}Information Systems Technology and Design Pillar, Singapore University of Technology and Design, Singapore\\
		Email: The corresponding author is Qin Hu (qinhu@iu.edu).
		}
	}
\maketitle
\pagestyle{empty}

\begin{abstract}
Metaverse as the next-generation Internet provides users with physical-virtual world interactions. To improve the quality of immersive experience, users access to Metaverse service providers (MSPs) and purchase bandwidth resource to reduce the communication latency of the Metaverse services. The MSPs decide selling price of the bandwidth resource to maximize the revenue. This leads to a joint  user association and resource pricing problem between all users and MSPs. To tackle the problem, we formulate a Stackelberg game where the MSPs are game leaders and users are game followers. We resolve the Stackelberg equilibrium via the distributed and centralized approaches, according to different privacy requirements. In the distributed approach, the MSPs compete against each other to maximize the individual revenue, and a user selects an MSP in a probabilistic manner. The Stackelberg equilibrium is achieved in a privacy-friendly way. In the centralized approach, all MSPs and users accept the unified management and their strategies are instructed. The centralized approach acquires the superior decision-making performance but sacrifices the privacy of the game players. Finally, we provide numerical results to demonstrate the effectiveness and efficiency of our schemes.
\end{abstract}

\begin{IEEEkeywords}
Metaverse, privacy, distributed, centralized and Stackelberg game.
\end{IEEEkeywords}

\section{Introduction}
Recently, Metaverse has attracted enormous attention due to the great potential application in permitting people to interact with each other and virtual objects in the three-dimensional (3D) virtual world, meanwhile providing them with immersive experience through digital avatars \cite{ning2021}. Advanced technologies such as virtual reality (VR), augmented reality (AR), 6G techniques and remote rendering, are integrated into the Metaverse to make the physical-virtual world interactions more real-time and seamless \cite{xu2022}. Nowadays, users with head-mounted displays (HMDs) can conveniently enjoy a variety of Metaverse applications, such as online multi-player game Minecraft, virtual art shows and concerts \cite{lee2021}.   Metaverse service providers (MSP) lease available resources (e.g., bandwidth, CPU and GPU) from the edge computing platforms, and provides reliable access services  and perform the remote rendering tasks for the users \cite{9612025}. We pay attention to the scenario where each user selects a proper MSP and  presents the bandwidth requirement to achieve a low-latency immersive experience while each MSP optimizes the selling price of bandwidth resource to attract the users and maximize the revenue. This leads to a joint user association and resource pricing problem between the users and MSPs, which is crucial for the wide implementation of Metaverse.

Research efforts have been conducted to improve the performance of Metaverse applications with different optimization objectives. For example, a VR service call market was studied in \cite{xu2021} to tackle the user matching and resource pricing problem between the VR users and service providers.  Double Dutch auction with deep reinforcement learning was proposed as the solution methodology. In \cite{du2021}, a new Meta-Immersion metric was introduced  to measure the quality of immersive experience of a user. Then the authors addressed the allocation problem of the basic bandwidth and acceleration bandwidth, according to different quality requirements of the access services. A comprehensive scheme based on the code distributed computing and collaborative blockchains was presented to accelerate the rendering task processing in parallel but also realize the decentralized management for Metaverse services \cite{jiang2021}. A game theory approach was further adopted to provide considerate rewards for the corresponding task performers. Similarly, the authors in \cite{kang2022} proposed a cross-chain empowered federated learning framework to provide security guarantee for data usage when integrating Metaverses into industrial IoT networks. In the virtual world systems (e.g., WAVE \cite{380746} and MASSIVE-3 \cite{greenhalgh2000}), a variety of virtual products are created  by different users. The user-generated contents need to be maintained well to guarantee the persistency of the virtual world \cite{shen2020}. To this end, an equity‑based incentive mechanism was designed to stimulate the users to collectively contribute the accessible computing resource and content storage for the virtual world contents. In addition, prototype implementation was conducted for Metaverse.  A blockchain-driven university Metaverse was developed to  perform the social experiments on the on-campus students  \cite{duan2021}. The Unity development platform was applied to investigate virtual education in the Metaverse and study the optimal resource allocation in the highly dynamic demand environment  \cite{ng2021}.  The potential technologies for edge-driven Metaverse was summarized in \cite{lim2022}, which also developed a virtual smart city as a case study.

However, there are challenging problems that need much further investigation for the wide deployment of Metaverse. On the MSP side, the decisions can be optimized via different approaches including distributed and centralized approaches. In the first approach,  all MSPs are selfish and compete against each other to obtain the individual revenue of selling the bandwidth resource. Due to the competitive effect, the MSP determines the  pricing strategies in a fully noncooperative manner. In this case, a user may access to one of the MSPs in a probabilistic manner. The probability is evaluated based on multiple factors, such as the selling price of the bandwidth resource and actual quality of communication link.  Note that the distributed approach permits each MSP to independently make a decision and keep the individual information private. In turn, the centralized approach is applicable for the scenario where all MSPs belong to a common management department such that they accept the unified management for the achievement of a certain goal. The strategies of the MSPs and users are instructed by a centralized decision maker. Then the joint user association and resource  pricing problem is tackled in a centralized manner. Thus, this requires each entity to share the individual information to the centralized decision maker, as a result, violating the privacy of the MSPs and users. Nowadays, the privacy awareness of users is enhanced in many interactive applications, particularly in online social networks \cite{9403974,9699411}. This motivates us to  study the comprehensive service optimizations for Metaverse according to different privacy requirements.

Based on the above considerations, we tackle the joint user association and resource pricing problem by using the distributed and centralized approaches, which are distinguished by the privacy policies. We first formulate the strategic interactions between the users and MSPs as a multi-leader multi-follower Stackelberg game. Each MSP is a game leader that sells the bandwidth resource to the users for maximizing the revenue while each user is a game follower that acts in response to the MSPs' strategies and decides the bandwidth sales for maximizing the utility. We resolve the Stackelberg game under different decision-making conditions. In the distributed approach, the competitive effect among the MSPs causes a noncooperative game in the upper stage. In the bottom stage, a user randomly selects an MSP according to a probability,  which is evaluated by a performance-price ratio of the MSP. We provide the theoretic analysis to prove the existence and uniqueness of Stackelberg equilibrium. Then the Stackelberg equilibrium is achieved without the collection of the private information. In the centralized approaches, the MSPs are instructed to design the pricing strategies and serve the given users for maximizing the total revenue of all MSPs.  The joint user association and resource pricing problem is formulated as a mixed integer nonlinear programming (MINP) problem after knowing private information of all game players. To seek the globally optimal solution, we develop a dynamic bound tightening algorithm. Compared with the distributed approach, the centralized approach obtains the superior decision-making performance but sacrifices the privacy of the game players.

The main contributions of the paper can be summarized as follows.
\begin{itemize}
	\item We present a system model to support the Metaverse services. Each MSP rents a dedicated edge server with available computing and communication resources to provide 3D video rendering and encoding while the users receive, decode and display the rendered video frames by using their HMDs.

	\item We formulate the Stackelberg game between the MSPs and users to study the joint optimization problem of user association and resource pricing for Metaverse. According to different application scenarios, each user is proactive or instructed to access to an MSP. The MSPs can independently determine the pricing strategies or follow the given instructions to serve the users.

	\item We derive the Stackelberg equilibrium under different privacy requirements. The distributed and centralized approaches are discussed when the MSPs provide the bandwidth sales for the users. In the approaches, we provide the theoretic analysis and efficient solutions to resolve the Stackelberg game.  
	
\end{itemize}

The rest of this paper is organized as follows. Section II presents the system model.
We introduce the Stackelberg game based service optimization and discuss it in the distributed and centralized environment in Sections III and IV, respectively.  Numerical results are shown in Section V. Finally, Section VI concludes this paper and discusses the future direction.

\begin{figure}
	\centering
	\includegraphics[width=0.45\textwidth]{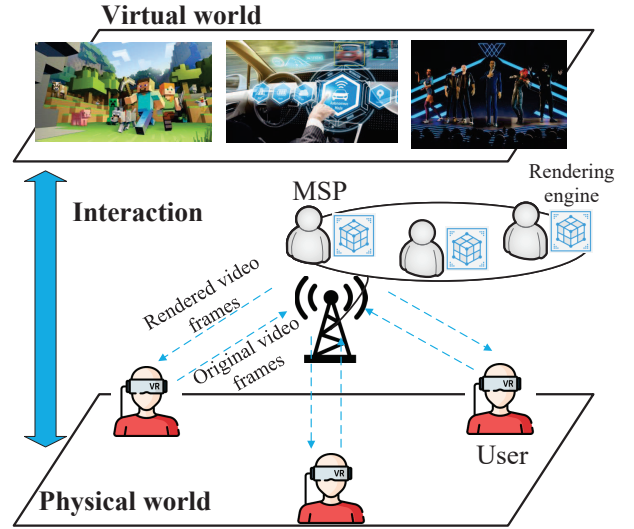}
	\caption{System model.}
	\label{s0}
\end{figure}

\section{System Model}
\subsection{Network Entity}
As shown in Fig.~\ref{s0}, edge assisted remote rendering is applied in the Metaverse to offload computation-intensive rendering tasks to proximal edge servers, and provide ultra-reliable and lower-latency access services for users. More specifically, a user with the HMD could run a VR application, and delivers the original video frames to an MSP through the nearest base station. The MSP starts an edge server that is associated with the local base station to serve the user. The edge server handles a real-time 3D rendering engine, which consists of hundreds or thousands of graphics processing unit cores, to render and encode the received video frames. After that, the rendered video frames are transmitted to the user that decodes and displays the video frames by the HMD.  To reinforce the immersive experience, the user requires sufficient bandwidth resource to enhance the data transmission and meet the delay requirement of the Metaverse application.  Then the user pays to the MSP according to the service level agreement. We provide more details of the system model as follows.

\begin{itemize}
	\item User: The widespread use of HMD allows the user to experience the virtual world and  interact with others and virtual objects in a variety of Metaverse applications, such as massively multiplayer online games, remote surgery, security applications in ITS, virtual art shows and concerts. To pursuit the high quality of the immersive experience, the user focuses on the interaction latency which consists of several parts: communication delay, rendering delay and video coding/decoding delay during the service provision. 
	
	\item MSP: The MSP provides access services for the users, and rents available bandwidth resource from the local wireless service provider through a long-term resource reservation plan. After being authorized, the MSP manages a number of communication channels, and specifies that a communication channel is assigned to a single user during the data transmission. The orthogonal frequency division multiple access technology is applied to ensure that all communication channels occupied by different users and MSPs are orthogonal, and thus inter-channel interference is negligible. For simplicity, we consider that the MSP uniformly allocates the processing capability of the rendering engine for all served users, and assigns different bandwidth sales to different users according to the service level agreements. The selling price of bandwidth resource of the MSPs is differential. In the distributed approach, the MSPs compete against each other and independently determine the pricing strategies while the MPs in the centralized approach are instructed by a centralized decision maker to present the pricing strategies and serve the given users.
	
	\item Metaverse service market: The Metaverse service market is periodically scheduled within several discrete time slots. A group of users and MSPs are confirmed in a scheduling period. When the MSPs are noncooperative, a user prefers to randomly select one of the MSPs, afterwards rationally determines the bandwidth sales to maximize the utility. In turn, an MSP adjusts the selling price of bandwidth resource to maximize the expected revenue. When all users and MSP are cooperative, access choices of the users and pricing strategies of the MSPs follow the instructions of the centralized decision maker.  Note that the orchestration of the Metaverse services is accomplished within a finite-time interval by promptly tackling the joint user association and resource allocation problem between all users and MSPs.
\end{itemize}

\subsection{Metaverse Service}
Referring to \cite{xu2021}, we mainly optimize the communication delay to achieve the low-latency immersive experience, and particularly pay attention to the optimization of downlink bandwidth. According to the individual requirement,  the user reserves a channel with available bandwidth from an MSP to receive the rendered data at the required bitrate. Take a user $i$ and MSP $j$ as an example and the bandwidth sales of user $i$ purchased from MSP $j$ is denoted as $s_i^j$. To guarantee the quality of the immersive experience,  there is a lower limit of $s_i^j$, which is denoted as $s_i^{\min}$ and utilized to satisfy the minimal bitrate requirement of the user. Given the bandwidth $s_i^j$, the downlink bitrate of user $i$ is calculated by
\begin{equation}
	\gamma _i^j = s_i^j{\log _2}\left( {1 + \frac{{\rho _j^{{\rm{tx}}}h_j^0d{{_i^j}^{ - \varepsilon }}}}{{{N_0}}}} \right), \label{dr}
\end{equation}
where $\rho_j, h_j^{0}, d_i^j, \varepsilon$ and $N_0$ represent the transmitter power, received power at the reference distance $d_0=1$ m, distance between the user and MSP,  path-loss exponent and noise power spectral density, respectively. We consider that the channel state information is known to the MSP.  To measure the minimal bitrate requirement of the user in the Metaverse, we take a non-panoramic VR application as an example. According to the previous work \cite{kelkkanen2020}, SSIM and VMAF are two commonly utilized quality estimators for evaluating the quality of the immersive experience in the non-panoramic VR application. The reference has provided the estimation method of the minimal bitrate requirement, with the given SSIM and VMAF values. When the SSIM value of user $i$ is required by ${\rm{SSI}}{{\rm{M}}_i}$, the minimal bitrate requirement is estimated by
\begin{equation}
	\gamma_{i,1}^{\min } = {\left( {\frac{{1 - {\rm{SSI}}{{\rm{M}}_i}}}{{{\kappa _1} + {\kappa _2}{v_i}}}} \right)^{ - \frac{1}{{{\kappa _3} + {\kappa _4}{v_i}}}}},
\end{equation}
where $v_i$ is rotation speed of the HMD, $\kappa_1,\kappa_2,\kappa_3$ and $\kappa_4$ are four presetting coefficients. When the VMAF value of user $i$ is required by ${\rm{VMA}}{{\rm{F}}_i}$, the minimal bitrate requirement is estimated by
\begin{equation}
	\gamma_{i,2}^{\min } = \frac{{{\rm{VMA}}{{\rm{F}}_i} - {\kappa _1} - {\kappa _2}{v_i}}}{{{\kappa _3} + {\kappa _4}{v_i}}}.
\end{equation}
Finally, the minimal bitrate requirement is expressed by  $\gamma_i^{\min } = \max (\gamma_{i,1}^{\min },\gamma_{i,2}^{\min })$ according to the  given ${\rm{SSI}}{{\rm{M}}_i}$ and ${{\rm{VMA}}{{\rm{F}}_i}}$.  We substitute $r_{i}^{\min}$ into Eqn.~(\ref{dr}) to calculate $s_i^{\min}$.

\section{Distributed Approach with Privacy Considerations}
Within a time window, a batch of service requests are gathered and the MSPs prepare to process them.  In the following, we employ a Stackelberg game to model the interactions between the users and MSPs. To facilitate the seamless immersive experience, the users play as game followers and purchase bandwidth resource from the MSPs, which play as game leaders and determine the selling price of bandwidth resource. The distributed approach enables the MSPs to independently determine the differential pricing strategy in a  distributed manner while maintaining the private information.
\subsection{Utility Formation in the Metaverse}
There are $I$ users and $J$ MSPs in the Metaverse service market. A user is indexed by $i, 1\le i\le I$ and an MSP is indexed by $j, 1\le j\le J$. The decision of user $i$ refers to the bandwidth sales purchase  from MSP $j$ and denoted as  $s_i^j$.  The decision of MSP $j$ refers to the selling price of the bandwidth resource for each user denoted as $p^j$. Motivated by the previous works \cite{xu2021privacy,9492053}, we consider that user $i$ accesses to one of the MSPs in a probabilistic manner.  From the viewpoint of  the user, the probability of accessing to MSP $j$ denoted as $\lambda_i^j$ is influenced by the selling price $p^j$ and quality of communication link between them $q^j$. Thus, we introduce a performance-price ratio $r^j=q^j/p^j$ and formulate the pairing probability between the user and MSP as follows  
\begin{equation}
	\lambda_i^j=\frac{{{r^j}}}{{\sum\limits_{j = 1}^J {{r^j}} }}.
\end{equation}
Sufficient bandwidth sales are helpful to reduce the communication delay of the user and reinforce the seamless experience of immersion of the user. With the increase of $s_i^j$, the user feels and experiences more immersive in the Metaverse applications. This consequently improves the user satisfaction when enjoying the physical-virtual world interactions.  Clearly, with the increase of $s_i^j$, the satisfaction degree increases. On the other hand, the degree improvement gradually decreases with the increase of $s_i^j$. To satisfy the function characteristics, we  adopt a function $S_i^j = {\alpha _i} s_i^j(2s_i^{\max } - s_i^j)$ to evaluate the satisfaction degree of the user with respect to $s_i^j$. Here, $\alpha_i$ is a user-centric parameter, which indicates the user sensitivity to the latency, and $s_i^{\max } $ is an upper limit of $s_i^j$. Thus,  the expected utility of user $i$ is  expressed as follows
\begin{equation}
	{U_i} = \sum\limits_{ j=1}^J {\lambda _i^j(S_i^j - {p^j}s_i^j)}.
\end{equation} 
The user aims to optimize the decision $s_i^j$ to maximize $U_i$.

Given the probable choices of all users $\{\lambda_i^j\}_{\forall i}$, MSP $j$ calculates the expected revenue as follows
\begin{equation}
	{R^j} = {p^j}\sum\limits_{i=1}^I {\lambda _i^js_i^j}.
\end{equation}
The MSP optimizes the decision $p^j\in [0, p_j^{\max}]$ to maximize $R^j$. Note that $R^j$ is also influenced by the strategies of the other MSPs, which are expressed by using a strategy set of all MSPs except MSP $i$, denoted as   ${\bf{p}}^{-j}=\left\{p^1,\cdots,p^{j-1},p^{j+1},\cdots,p^{J}\right\}$. Since the MSPs compete against each other to serve the users to earn the revenue,  there is a noncooperative game among the MSPs.

\subsection{Stackelberg Equilibrium Analysis}
To obtain the solution of the formulated game, we seek the Stackelberg equilibrium, at which each MSP plays a game leader and maximizes the expected utility given the best responses of the users that play as the game followers. At this equilibrium, neither any game leader nor any game follower can improve the individual utility by unilaterally changing the strategy. We define the Stackelberg equilibrium as follows.

\textbf{Definition 1 (Stackelberg Equilibrium).} User $i$ and MSP $j$ finally obtains an optimal strategy vector ${{\bf{s}}_i^*}=\{ s_i^{j*}\} _{j = 1}^J$ and an optimal selling price of the bandwidth resource $p^{j*}$, respectively. We say that a set of strategies $({{\bf{s}}^*} = {\left\{ {{\bf{s}}_i^*} \right\}_{i=1}^{I}},{{\bf{r}}^*} = {\left\{ {{r^{j*}}} \right\}_{j=1}^{J}})$  can be the Stackelberg equilibrium if and only if the following set of inequalities is strictly satisfied:
\begin{equation*}
\left\{ {\begin{array}{*{20}{l}}
		{{R^j}({p^{j*}},{{\bf{p}}^{ - j*}},{{\bf{s}}^*}) \ge {R^j}({p^j},{{\bf{p}}^{ - j*}},{{\bf{s}}^*}),\forall j,}\\
		{{U_i}({\bf{s}}_i^*,{{\bf{p}}^*}) \ge {U_i}({{\bf{s}}_i},{{\bf{p}}^*}),\forall i.}
\end{array}} \right.
\end{equation*}

In the following, we adopt the backward induction method to discuss the Stackelberg equilibrium. To simplify the theoretic analysis, we temporarily neglect the lower limit of $s_i^j$ and consider an infinite bandwidth capacity for each MSP. We derive the best response of a game follower by taking the first and second order derivatives of $U_i$ with respect to $s_i^j$,
\begin{equation*}
\begin{array}{*{20}{l}}
	{\frac{{\partial {U_i}}}{{\partial s_i^j}} = \lambda _i^j(2{\alpha _i}s_i^{\max } - 2{\alpha _i}s_i^j - {p^j}),}\\
	{\frac{{{\partial ^2}{U_i}}}{{\partial s_i^{j2}}} =  - 2\lambda _i^j{\alpha _i}s_i^j < 0.}
\end{array}
\end{equation*}
The negative second-order derivative of $U_i$ indicates that $U_i$ is strictly concave with respect to $s_i^j$. Based on the  first-order optimality condition, we express the best response $s_i^{j*}$ by
\begin{equation}
s_i^{j*} = s_i^{\max } - \frac{{{p^j}}}{{2{\alpha _i}}},
\end{equation}
We always have $s_i^{j*}<s_i^{\max}$ and with the increase of $p^j$, less bandwidth sales are required by the user.  Considering the nonnegative value of $s_i^j$, we further have 
\begin{equation}
	s_i^{j*}=\max (s_i^{\max } - \frac{{{p^j}}}{{2{\alpha _i}}},0).\label{opts}
\end{equation}

The MSPs are selfish to serve the users and obtain the revenue in a competitive manner. This results in a noncooperative game among them, where the player set including $J$ MSPs is finite and the strategy spaces of all MSPs are nonempty, convex, and compact subsets of the Euclidean spaces. The revenue function is also continuous in the strategy space. To analyze the existence of the Nash equilibrium of the noncooperative game, we study the concavity of the utility function of any MSP. Substituting $s_i^{j*}$ into ${V}^j$, we have
\begin{equation}
{R^j} = \frac{{{q^j}}}{{\sum\limits_{j=1}^J {\frac{{{q^j}}}{{{p^j}}}} }}\sum\limits_{i=1}^I {\left( {s_i^{\max } - \frac{{{p^j}}}{{2{\alpha _i}}}} \right)} =\frac{{{X^j} - {p^j}{Y^j}}}{{\sum\limits_{j=1}^{J} {\frac{{{q^j}}}{{{p^j}}}} }},
\label{fR}
\end{equation}
where ${X^j} = {q^j}\sum\nolimits_{i=1}^I {s_i^{\max }} ,{Y^j} = {q^j}\sum\nolimits_{i=1}^{I} {\frac{1}{{2{\alpha _i}}}}$.  Since $s_i^{j*}\ge 0, \forall i, j$, we get that $X^j-p^jY^j \ge 0$ and $R^j \ge 0$. We take the first  and second order derivatives of ${R}^j$ with respect to $p^j$, 
\begin{equation*}
		\begin{array}{l}
			\frac{{\partial{R^j}}}{{\partial {p^j}}} = \frac{{ - {Y^j}\sum\limits_{k \ne j} {\frac{{{q^k}}}{{{p^k}}}}  + \frac{{{q^j}{X^j}}}{{{p^{j2}}}} - 2\frac{{{q^j}{Y^j}}}{{{p^j}}}}}{{{{\left( {\sum\limits_{ j=1}^J {\frac{{{q^j}}}{{{p^j}}}} } \right)}^2}}} = \frac{{{\pi ^j}}}{{{{\left( {\sum\limits_{j=1}^J {\frac{{{q^j}}}{{{p^j}}}} } \right)}^2}}},\\
			\frac{{{\partial ^2}{R^j}}}{{\partial {p^{j2}}}} = 2\frac{{{\pi ^j}}}{{{{\left( {\sum\limits_{j=1}^J {\frac{{{q^j}}}{{{p^j}}}} } \right)}^3}}}\frac{{{q^j}}}{{{p^{j2}}}} - \frac{{({X^j} - {p^j}{Y^j})}}{{{{\left( {\sum\limits_{j=1}^J {\frac{{{q^j}}}{{{p^j}}}} } \right)}^2}}}\frac{{2{q^j}}}{{{p^{j3}}}}.
		\end{array}
\end{equation*}
If $\partial{{R}^j}/\partial{p^j}=0$, $\pi^j =0$ and ${\partial ^2}{R^j}/{\partial ^2}{p^{j2}}<0$ is satisfied. Thus, utility function of MSP $j$ is quasi-concave and we have proved that the noncooperative game among the MSPs admits a Nash equilibrium.

Next, we prove the uniqueness of the Nash equilibrium by studying the best response function of any MSP. According to the first-order optimality condition $\partial {R^j}/\partial {p^j}=0$, we express the best response of MSP $j$, 
\begin{equation}
	{p^{j*}} = \left\{ \begin{array}{l}
		{F^j}({\bf{p}}),~0 \le {{\cal{F}}^j}({\bf{p}}) < p_j^{\max },\\
		p_j^{\max },~{{\cal{F}}^j}({\bf{p}}) \ge p_j^{\max }.
	\end{array} \right.
\end{equation}
where
\begin{equation}
	{\mathcal{F}}^j({\bf{p}}) =\frac{{2{q^j}{X^j}}}{{2{q^j}{Y^j} + \sqrt {{{\left( {2{q^j}{Y^j}} \right)}^2} + {{4{q^j}{X^j}{Y^j}}}\sum\limits_{k \ne j} {\frac{{{q^k}}}{{{p^k}}}} } }}.
\end{equation}
If $p^{j*}<p_j^{\max}$, we judge whether ${\cal F}^j ({\bf{p}})$ is a standard function, which should satisfy the following  properties:
\begin{itemize}
	\item \textit{Positivity:} ${{\cal{F}}^j(\bf{p})} >0$.
	\item \textit{Monotonicity:} if $\bf{p}^{\prime}>\bf{p}$, ${\cal F}^j({\bf{p}^{\prime}}) > {\cal F}^j({\bf{p}})$.
	\item \textit{Scalability:} $\forall \beta>1$, $\beta{\cal F}^j({\bf{p}}) > {\cal F}^j({\beta\bf{p}})$.
\end{itemize}
First, when $q^j >0, s_i^{\max}>0, \alpha_i>0, \forall i, j$, ${\cal{F}}^j(\bf{p}) >0$. Second, if $\bf{p}^{\prime}>\bf{p}$, we have $p^{j\prime}>p^{j}, \forall j$ and $\sum\nolimits_{k \ne j} {\frac{{{q^k}}}{{{p^{k\prime }}}}}  < \sum\nolimits_{k \ne j} {\frac{{{q^k}}}{{{p^k}}}}$. We further have  ${\cal{F}}^j({\bf{p}^{\prime}}) > {\cal{F}}^j({\bf{p}})$. Third, we explain that since $\beta >1$, $\beta^2 > \beta$
\begin{equation*}
	\begin{aligned}
	{{\mathcal F}^j}(\beta {\bf{p}}) &= \frac{{2{q^j}{X^j}}}{{2{q^j}{Y^j} + \sqrt {{{\left( {2{q^j}{Y^j}} \right)}^2} + \frac{{4{q^j}{X^j}{Y^j}}}{\beta }\sum\limits_{k \ne j} {\frac{{{q^k}}}{{{p^k}}}} } }}\\
	&< \frac{{2{q^j}{X^j}}}{{2{q^j}{Y^j} + \sqrt {{{\left( {\frac{{2{q^j}{Y^j}}}{\beta }} \right)}^2} + \frac{{4{q^j}{X^j}{Y^j}}}{{{\beta ^2}}}\sum\limits_{k \ne j} {\frac{{{q^k}}}{{{p^k}}}} } }}\\
	& < \frac{{2{q^j}{X^j}}}{{\frac{{2{q^j}{Y^j}}}{\beta } + \sqrt {{{\left( {\frac{{2{q^j}{Y^j}}}{\beta }} \right)}^2} + \frac{{4{q^j}{X^j}{Y^j}}}{{{\beta ^2}}}\sum\limits_{k \ne j} {\frac{{{q^k}}}{{{p^k}}}} } }} = \beta {{\mathcal F}^j}({\bf{p}}).
	\end{aligned}
\end{equation*}
If $p^{j*}=p_j^{\max}$, the best response function ${\cal F}^j ({\bf{p}})$ is still a standard function. We have proved that the best response function of any MSP is the standard function. According to the prior knowledge in \cite{han2012}, the Nash equilibrium of the non-cooperative game  is unique. Until now, we know that the best response of any game follower is unique  and the Nash equilibrium among the leaders is also unique. Finally, we conclude that there exists a unique Stackelberg equilibrium in the proposed Stackelberg game.

After proving the existence and uniqueness of the Stackelberg equilibrium,  we apply the conventional best-response dynamics algorithm to reach the Stackelberg equilibrium. With the given responses of the users, we utilize the following rule to update the strategy of each MSP in each iteration $k$
\begin{equation}
	p^j_k = p^j_k + \mu p^j_{k-1}\frac{{\partial {{R}^j}}}{{\partial {p^j}}},~\forall j,
\end{equation}
where $\mu$ is a learning rate,  and  ${\partial {{R}^j}}/{{\partial {p^j}}}$ is approximately calculated by using the central difference method,
\begin{equation}
\frac{{\partial {R^j}}}{{\partial {p^j}}} \approx \frac{{R_{k + }^j - R_{k - }^j}}{{2\Delta p}},
\end{equation}
where $R_{k + }^j = {R^j}(p_{k - 1}^j + \Delta p,{\bf{p}}_{k - 1}^{ - j})$ and $R_{k - }^j = {R^j}(p_{k - 1}^j - \Delta p,{\bf{p}}_{k - 1}^{ - j})$. Let all MSPs update the strategies in parallel in the algorithm. Note that each MSP and user in each iteration are only required to feedback the dynamic responses and not necessitated to reveal the private information. Ultimately, the Stackelberg equilibrium is derived in a distributed and privacy-friendly manner.
\section{Centralized Approach without Privacy Protection}
The above Stackelberg game model is still adopted. Alternatively, we  pay attention to the centralized service optimization in the Metaverse, where all MSPs and users accept the unified management and follow the instructions from a centralized decision maker to take the actions. Furthermore, we considering feasible constraints for both the users and MSPs, and formulate the joint user connection and resource pricing problem as a MINLP problem. After that, we develop a dynamic bound tightening algorithm for obtaining the globally optimal solution to the MIMLP problem. To guarantee the superior decision-making performance, private information of all game players is collected as the prior knowledge of the centralized decision maker.

\subsection{Problem Formulation}
The centralized decision maker defines $x_i^j$ as a binary variable indicating the association between a user and an MSP. Note that $x_i^j=1$ means that user $i$ is served by MSP $j$, $x_i^j = 0$ otherwise. Each user can be served by at most one MSP, and we enforce
\begin{align}
\sum_{j=1}^J
 x^j_i & \le 1,
\forall i,
\label{eq_x_1}
\\
x_i^j & \in \{ 0, 1 \}, \forall i,j. 
\label{eq_x_01}
\end{align}
The upper limit $p^{\max}_j$ is considered for the price $p_j$,
\begin{align}
0 \le p^j \le p^{\max}_j,
\forall j. 
\label{eq_p_bound}
\end{align}
The bandwidth sales of user $i$ purchased from MSP $j$ are restricted by
\begin{align}
s^{\min}_i x^j_i
\le s^j_i
\le s^{\max}_i x^j_i,
\forall i,j.
\label{eq_s_bound}
\end{align}
It can be seen that if user $i$ purchases bandwidth resource from MSP $j$, i.e., $x^j_i = 1$, we have
$s^{\min}_i \le s^j_i \le s^{\max}_i$.
If user $i$ is not served by MSP $j$, i.e., $x^j_i = 0$, the bandwidth sales are $s^j_i = 0$. The bandwidth sales of user $i$ satisfy
\begin{align}
s^j_i =
\Big( s^{\max}_i
- \frac{p^j}{2 \alpha_i}
\Big) x^j_i,
\forall i,j,
\label{eq_s_p_x}
\end{align}
which also ensures $s^j_i = 0$ if $x^j_i = 0$.  As a result, the decision of user $i$ is constrained by 	$s_i^{\min} \le s_i^j  \le s_i^{\max }$.

According to the application scenario, we consider a finite bandwidth capacity $C^j$ for MSP $j$,
\begin{equation}
\sum_{i=1}^I
s_i^j \le {C^j},
\forall j.
\label{eq_C}
\end{equation} 
The objective function of the centralized decision maker is to maximize the weighted sum of the revenue of the MSPs. Here, a weighting factor of MSP $j$ is ${\varpi ^j} = {q^j}/\sum\nolimits_{j = 1}^J {{q^j}}$, which ensures that with the higher quality of the communication link, a higher price is normally required.  The optimization problem with necessary constraints is shown  as follows.
\begin{align}
\max_{\bm{p},\bm{s},\bm{x}}
&
\sum_{i=1}^I
\sum_{j=1}^J
\varpi ^j p^j s_i^j,
\label{eq_cooperative_obj}
\\
\text{s.t. } &
\text{(\ref{eq_x_1})--(\ref{eq_C})},
\nonumber
\end{align}
where the decision variables are
$\bm{p} = \{ p^j \}_{j=1}^{J}$,
$\bm{s} = \{\{ s^j_i \}_{j=1}^{J} \}_{i=1}^{I}$, and
$\bm{x} = \{\{ x^j_i \}_{j=1}^{J} \}_{i=1}^{I}$.
The problem is a typical MILP problem, where the nonconvexity comes from the bilinear terms $p^j s_i^j$ in objective (\ref{eq_cooperative_obj}) and
$p^j x_i^j$ in constraint (\ref{eq_s_p_x}).
Note that in this problem, user $i$ may receive zero bandwidth, i.e., $\sum_{j=1}^J x^j_i = 0$ and $\sum_{j=1}^J s^j_i = 0$.
If user $i$ has $x^j_i = 1$ for MSP $j$, the received bandwidth is required to satisfy $s^j_i \ge s^{\min}_i$ and ensure the quality of immersive experience. To tackle the MINLP problem, we first linearize the original problem to make it more tractable.

\subsection{Problem Linearization}
We use McCormick envelope \cite{McCormick_1976} to relax the bilinear terms in problem (\ref{eq_cooperative_obj}). Based on (\ref{eq_x_01}) and (\ref{eq_p_bound}), we define the following linear constraints for $p^j x_i^j$, which compose a convex envelope.
\begin{subequations}
\label{eq_Mc_px}
\begin{align}
{p^j x_i^j} 
&\ge 0,
\forall i, j,
\\
{p^j x_i^j} 
&\ge p^{\max}_j x^j_i + p^j - p^{\max}_j,
\forall i, j,
\\
{p^j x_i^j} 
&\le p^j,
\forall i, j,
\\
{p^j x_i^j} 
&\le p^{\max}_j x^j_i,
\forall i, j,
\end{align}
\end{subequations}
where we consider ${p^j x_i^j}$ as a new individual variable.
It can be seen that ${p^j x_i^j} = 0$ if $x_i^j = 0$, and ${p^j x_i^j} = p^j$ if $x_i^j = 1$.
This means that (\ref{eq_Mc_px}) is an exact relaxation for bilinear term $p^j x_i^j$.
Thus, bilinear term $p^j x_i^j$ can be linearized by simply taking $p^j x_i^j$ as a variable and adding constraints (\ref{eq_Mc_px}) to problem (\ref{eq_cooperative_obj}).
Similarly, we can define the convex envelope for bilinear term $p^j s_i^j$ as follows. 
\begin{subequations}
\label{eq_Mc_ps}
\begin{align}
{p^j s_i^j} 
&\ge p^{\min}_j s^j_i,
\forall i, j,
\\
{p^j s_i^j} 
&\ge p^{\max}_j s^j_i + p^j s^{\max}_i - p^{\max}_j s^{\max}_i,
\forall i, j,
\\
{p^j s_i^j}  
&\le p^{\min}_j s^j_i + p^j s^{\max}_i
- p^{\min}_j s^{\max}_i,
\forall i, j,
\\
{p^j s_i^j} 
&\le p^{\max}_j s^j_i,
\forall i, j,
\end{align}
\end{subequations}
where $p^j s_i^j$ is also taken as an individual variable.
Inequalities (\ref{eq_Mc_ps}) are derived based on
$p^{\min}_j \le p^j \le p^{\max}_j$
and $0 \le s^j_i \le s^{\max}_i$.
We have
$p^{\min}_j = 0$ in (\ref{eq_p_bound}), but we explicitly write $p^{\min}_j$ in (\ref{eq_Mc_ps}) for 
ease of explanation.
It can be seen that relaxation (\ref{eq_Mc_ps}) is inexact for bilinear term $p^j s_i^j$.
If relaxation (\ref{eq_Mc_ps}) is applied to problem (\ref{eq_cooperative_obj}), then the resulting relaxed problem may have a higher optimal objective value than problem (\ref{eq_cooperative_obj}).

To reduce the error caused by relaxation (\ref{eq_Mc_ps}), we can partition the domains of continuous variables involved in the bilinear terms \cite{Bergamini_2008, Nagarajan_2016, Zhong2020cooperative}.
The idea is to divide the domain of $p^j$ into a number of partitions and enforce that only one of these partitions is activated.
Then, we apply relaxation (\ref{eq_Mc_ps}) to the activated partition rather than the entire domain $[0, p^{\max}_j]$.
To divide the domain of $p^j$, $[0, p^{\max}_j]$, into $N_j$ partitions, we define a set of points $\bm{P}^j = \{P^j_k\}^{N_j}_{k=0}$, where
$P^j_k < P^j_{k+1}$,
$P^j_0 = 0$ and
$P^j_{N_j} = p^{\max}_j$.
For each partition $k \in \{1,\ldots,N_j\}$, define a lower bound
$L^j_k = P^j_{k-1}$, an upper bound
$U^j_k = P^j_{k}$, and a binary variable $y^j_k$ indicating that partition $k$ is activated or not.
We have
\begin{align}
\sum_{k=1}^{N_j}
 y^j_k & = 1,
\forall j,
\label{eq_y_1}
\\
y_k^j & \in \{ 0, 1 \}, \forall j,k.
\label{eq_y_01}
\end{align}
Thus, the lower and upper limits of the activated partition of $p^j$ are given by
\begin{align}
L^j_a & = 
\sum_{k=1}^{N_j}
 L^j_k y^j_k,
\forall j,
\label{eq_La}
\\
U^j_a & = 
\sum_{k=1}^{N_j}
 U^j_k y^j_k,
\forall j.
\label{eq_Ua}
\end{align}
For instance, if partition $\ell$ is activated for $p^j$, then
$y^j_{\ell} = 1$ and
$L^j_a = L^j_{\ell} \le p^j
\le U^j_{\ell} = U^j_a$ hold.
This indicates that we always have
$L^j_a \le p^j \le U^j_a$, based on which the McCormick envelope (\ref{eq_Mc_px}) for bilinear term $p^j s^j_i$ is provided by 
\begin{subequations}
\label{eq_Mc_ps_active}
\begin{align}
{p^j s_i^j} 
&\ge L^j_a s^j_i,
\forall i, j,
\\
{p^j s_i^j} 
&\ge U^j_a s^j_i + p^j s^{\max}_i - U^j_a s^{\max}_i,
\forall i, j,
\\
{p^j s_i^j} 
&\le L^j_a s^j_i + p^j s^{\max}_i
- L^j_a s^{\max}_i,
\forall i, j,
\\
{p^j s_i^j} 
&\le U^j_a s^j_i,
\forall i, j.
\end{align}
\end{subequations}
Constraints (\ref{eq_Mc_ps_active}) also contain bilinear term $y^j_k s^j_i$, which has a binary variable in the product.
Similar to (\ref{eq_Mc_px}), McCormick relaxation is exact for $y^j_k s^j_i$.
Thus, (\ref{eq_Mc_ps_active}) can be linearized exactly.

Finally, the linearized version of problem (\ref{eq_cooperative_obj}) is given by
\begin{align}
\max_{\substack{\bm{p},\bm{s},\bm{x}, {\bm{px}},
\\  \bm{y}, {\bm{ps}}, {\bm{ys}} }}
&
\sum_{i=1}^I
\sum_{j=1}^J
{\varpi ^j p^j s_i^j},
\label{eq_linearized_obj}
\\
\text{s.t. } &
\text{(\ref{eq_x_1})--(\ref{eq_C}),
(\ref{eq_Mc_px}),
(\ref{eq_y_1})--(\ref{eq_Mc_ps_active})},
\nonumber
\\
& \text{McCormick envelope for }
y^j_k s^j_i,
\forall,i,j,k,
\nonumber
\end{align}
where $p^j x^j_i$, $p^j s^j_i$, and
$y^j_k s^j_i$ are individual variables, and we define
$\bm{px} = \{\{ p^j x^j_i \}_{i=1}^I\}_{j=1}^j$,
$\bm{ps} = \{\{ p^j s^j_i \}_{i=1}^I\}_{j=1}^j$,
$\bm{ys} = \{\{ y^j_k s^j_i \}_{i=1}^I\}_{j=1}^J\}_{k=1}^{N_j}$, and
$\bm{y} = \{\{ y^j_k \}_{j=1}^J\}_{k=1}^{N_j}$.
This is an MILP problem, which can be solved by off-the-shelf solver, e.g., CPLEX.
Problem (\ref{eq_linearized_obj}) is still a relaxation of problem (\ref{eq_cooperative_obj}), but the relaxation error is reduced via partitioning.
If the sizes of the activated partitions (i.e., $U^j_a - L^j_a$) are small enough, problem (\ref{eq_linearized_obj}) can be approximately equivalent to problem (\ref{eq_cooperative_obj}).
As shown by (\ref{eq_La}) and (\ref{eq_Ua}), the selection of $L^j_k$ and $U^j_k$ (i.e., $\bm{P}^j$) is significantly important for the performance of such approximation.
A simple method is to evenly partition $[0, p^{\max}_j]$ with small partition sizes.
This can guarantee a small $(U^j_a - L^j_a)$ but leads to a lot of binary variables (i.e., $\bm{y}$), which considerably expand the size to problem (\ref{eq_linearized_obj}).
Next, we develop an algorithm to dynamically generate $\bm{P}^j$, using a reduced number of binary variables to achieve high-quality approximation.

\begin{algorithm}[t]
\caption{Bound tightening algorithm} 
\label{algorithm1}
\small
Input: $\beta > 1$,
$\epsilon > 0$,
$UB = + \infty$,
$LB = - \infty$,
$\bm{P}^j = \{ 0, p^{\max}_j \}, \forall j$.
\\
Output: Globally optimal solution to problem (\ref{eq_cooperative_obj}).
\\
\Repeat
{$ LB = UB $}
{
Given $\bm{P}^j,\forall j$, solve the linearized problem (\ref{eq_linearized_obj}) and get the optimal solution
$\{ \bm{p}^*,  \bm{s}^*, \bm{ps}^*, \bm{y}^*\}$.
\\
\For{$j=1,\ldots,J$}
{Let $\ell$ be the index of the activated partiton of $p^j$, that is
$y^{j*}_{\ell} = 1$.
\\
Define $Z^j_{new} = (U^j_{\ell} - L^j_{\ell})/{\beta}$.
\\
\eIf{$Z^j_{new} \le \epsilon$}
{No new partition is needed for $p^j$.
} 
{$L^j_{new} = \max\{ L^j_{\ell}, p^{j*} - Z^j_{new} \}$.
\\
$U^j_{new} = \min\{ U^j_{\ell}, p^{j*} + Z^j_{new} \}$.
\\
$\bm{P}^j \leftarrow \bm{P}^j \cup \{ L^j_{new}, U^j_{new} \}$.
} 
} 
$LB = \max\{ LB, \sum_{i=1}^I \sum_{j=1}^J p^{j*} s^{j*}_i \}$.
\\
$UB = \min\{ UB, \sum_{i=1}^I \sum_{j=1}^J p^{j} s^{j*}_i \}$.
} 
\end{algorithm}

\subsection{Bound Tightening Algorithm}
The proposed bound tightening algorithm is performed when private information of all game players is known by the centralized decision maker. More details of the algorithm are presented in Algorithm~\ref{algorithm1}.
The algorithm starts with $\bm{P}^j = \{ 0, p^{\max}_j \}, \forall j$, i.e., only one partition for $p^j$.
Given this $\bm{P}^j$, we get the optimal solution to the relaxed problem (\ref{eq_linearized_obj}).
Based on the solution, we tighten the bounds of the activated partitions.
In line~7 of Algorithm~\ref{algorithm1}, $\beta$ is a parameter for adjusting the size of new partitions.
In line~8, $\epsilon$ is a small positive number for determining whether a new partition is needed.
In lines~11--13, $L^j_{new}$ and $U^j_{new}$ are lower and upper bounds of a new partition.
After the point set $\bm{P}^j$ is updated, problem (\ref{eq_linearized_obj}) is solved in the next iteration.
The optimal objective value of problem (\ref{eq_linearized_obj}) is given by
$ \sum_{i=1}^I \sum_{j=1}^J p^{j} s^{j*}_i $, which provides the upper bound (UB) of the optimal objective value of the original problem (\ref{eq_cooperative_obj}).
Plugging $\{ \bm{p}^*, \bm{s}^* \}$ into the objective in (\ref{eq_cooperative_obj}) yields $ \sum_{i=1}^I \sum_{j=1}^J p^{j*} s^{j*}_i $, which is the lower bound (LB) of the optimal objective value of problem (\ref{eq_cooperative_obj}).
At each iteration, the relaxation error caused by (\ref{eq_Mc_ps_active}) may be reduced by tightening the bounds of active partitions.
Therefore, we eventually will have $LB = UB$.
When $LB = UB$, $\{ \bm{p}^*, \bm{s}^* \}$ is the globally optimal solution to the original problem (\ref{eq_cooperative_obj}).

\section{Numerical Results}
For evaluating the performance of the above schemes, we present simulation results in this section based on the following parameter setting.  There are $I=10$ users and $J=3$ candidate MSPs. For simplicity, their parameters follow the uniform distribution. On the user side, the user-centric parameter $\alpha \in U(0,1)$, the minimal and maximal bandwidth sales $s^{\min} \in U(1, 5)$ and $s^{\max}\in U(10, 12)$. On the MSP side, the quality of communication link $q\in U(0,1)$, the bandwidth capacity of three MSPs is set by $[20, 30, 50]$ MHz, and the upper limit of selling price of bandwidth resource is identical for each MSP $p^{\max}=12$.  In the best-response dynamics algorithm, step size $\Delta p$=1e-4 and learning rate $\mu$=1e-2. In the bound tightening algorithm, parameters $\beta=10$ and $\epsilon$=1e-3.

\begin{figure}[!t]
	\centering
	\subfigure[Convergence of the best-response dynamics algorithm]{
		\label{s1a} 
		\includegraphics[width=0.45\textwidth]{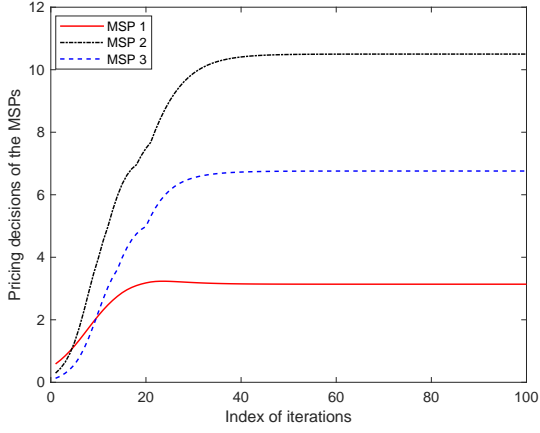}}
	\subfigure[Convergence of the bound tightening algorithm]{
		\label{s1b} 
		\includegraphics[width=0.45\textwidth]{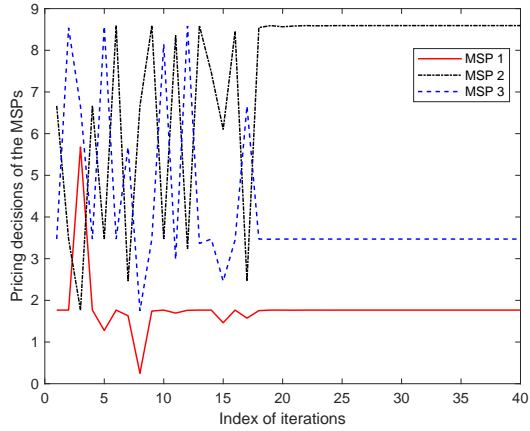}}
	\caption{Convergence of the proposed algorithms for the distributed and centralized schemes.}
	\label{s1}
\end{figure}

\begin{figure}[t!]
	\centering
	\includegraphics[width=0.45\textwidth]{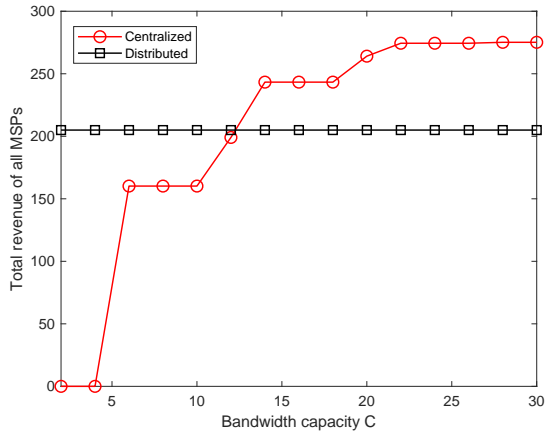}
	\caption{Comparison of the total revenue of all MSPs under different schemes with respect to different $C$.}
	\label{s2}
\end{figure}

The overall performance of the proposed algorithms designed for the distributed and centralized schemes is shown in Fig.~\ref{s1}. The changing strategies of the MSPs in different iterations are observed. Both the best-response dynamics algorithm and  bound tightening algorithm can finally converge within dozens of iterations. In addition, we conduct the performance comparison between the two schemes, as shown in Fig.~\ref{s2}.  Here, the bandwidth capacity of three MSPs is identical and denoted by $C$. The total revenue of all MSPs under different schemes with respect to $C$ is observed. From the figure, we know that when $C$ is small in the centralized scheme, no users with bandwidth demand can be served. With the increase of $C$, each MSP can serve more users and provide more bandwidth sales to improve the individual revenue. When $C$ becomes large enough, constraint (\ref{eq_C}) of each MSP is invalid and optimal strategy of each user is directly expressed by Eqn.~(\ref{opts}). Substituting Eqn.~(\ref{opts}) into the objective function in (\ref{eq_cooperative_obj}), the pricing strategies of the MSPs are purely optimized to maximize the reformulated objective function. Then the total revenue of all MSPs will become changeless. The total revenue of all MSPs in the distributed scheme is constant since the infinite bandwidth capacity is assumed for each MSP. Moreover, we observe that with the increase of $C$, the total revenue of all MSPs in the centralized scheme can successfully exceed that in the distributed scheme. Under the same condition on the bandwidth capacity, the centralized scheme is better than the distributed scheme in improving the total revenue of all MSPs. But the centralized scheme sacrifices the privacy of all game players due to the collection of the private information.

\begin{figure}[!t]
	\centering
	\subfigure[Comparison of the decision of the user with respect to different $\alpha$ and $p$]{
		\label{s3a} 
		\includegraphics[width=0.45\textwidth]{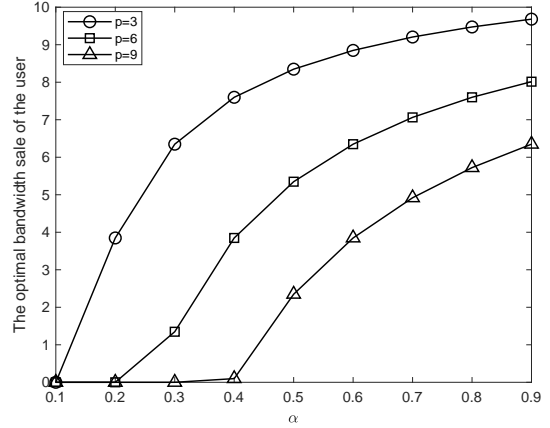}}
	\subfigure[Comparison of the decision of the MSP with respect to different $\alpha$ and $q$]{
		\label{s3b} 
		\includegraphics[width=0.45\textwidth]{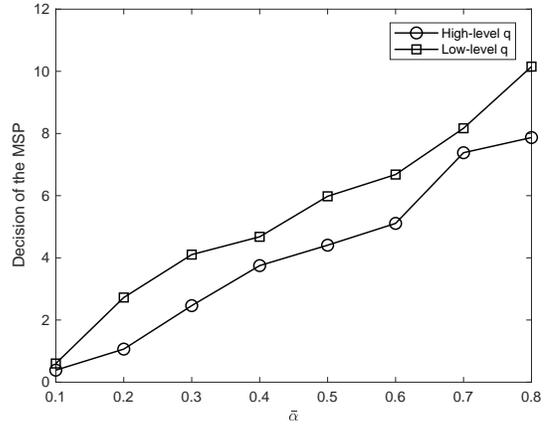}}
	\caption{Impacts of key parameters on the decisions of the game players.}
	\label{s3}
\end{figure}

Next, we investigate the impacts of key parameters such as user-centric parameter $\alpha$, selling price of bandwidth $p$ and quality of communication link $q$ on the optimal decisions of a user and an MSP. We take the first user and first MSP as an example. According to Eqn.~(\ref{opts}),  the decision of the user is in terms of the bandwidth sales purchased from the MSP, which is mainly influenced by internal and external factors, i.e., $\alpha$ and $p$. The $\alpha$ indicates the user sensitivity to the latency. With the increase of $\alpha$, more bandwidth is preferably necessitated to improve the quality of the immersion experience. In turn, when $\alpha$ is too small, the user sharply reduces the bandwidth sales, and to zero when necessary.  With the increase of $p$, monetary cost of getting the same bandwidth sales increases and the user naturally reduces the bandwidth sales. This is consistent with the intuition, as shown in Fig.~\ref{s3a}. For example, when $p$ increases from 3 to 9, the bandwidth sales of the user are decreased more than 70\% when $\alpha=0.5$. 

We also observe the changing decision of the MSP with varying $\alpha$ and $q$ in the centralized scheme.  Let $\bar{\alpha}$ denote the mean value of all $\alpha$. With the increase of $\bar{\alpha}$, most of the users are willing to buy more bandwidth resource and the MSP can increase the pricing strategy. We divide the quality of the communication link into two kinds, including high-level (i.e., $q=0.7$) and low-level (i.e., $q=0.3$).  Given the high-level $q$, the pricing strategy of the MSP is roughly improved by the centralized decision maker according to the objective function in (\ref{eq_cooperative_obj}).  With the higher quality of the communication link, the higher selling price of the bandwidth resource is normally required in Fig.~\ref{s3b}. For example, the selling price of the MSP increases about 31\% when $\bar{\alpha}=0.6$.

\section{Conclusion}
In this paper, we studied the service optimization for Metaverse via the distributed and centralized approaches, which were adopted to satisfy different privacy requirements. To feel and experience seamlessly immersive in the Metaverse services, different users purchase bandwidth resource from different MSPs. We applied a multi-leader multi-follower Stackelberg game to investigate the joint user association and resource pricing problem between all users and MSPs. The Stackelberg equilibrium was discussed by theoretic analysis, and derived by using efficient algorithms according to the competitive effect and cooperative awareness among the MSPs. Numerical results were provided to compare the distributed and centralized approaches, and to demonstrate the effectiveness and efficiency of our schemes. 

In the future, we will improve the mathematical model to adapt to the Metaverse services, and may develop a prototype system to evaluate our scheme.  In addition, we could enhance the solution methodology by using the AI tools, such as  deep reinforcement learning, to tackle the service optimization procedure over multiple time slots.


\section*{Acknowledgement}
This work was supported in part by the National Natural Science Foundation of China under Grant 62001125, Grant 62003099, and Grant 62102099, in part by the Open Research Project of the State Key Laboratory of Industrial Control Technology, Zhejiang University, China under Grant ICT2022B12, and in part by the National Research Foundation, Singapore and Infocomm Media Development Authority under its Future Communications Research \& Development Programme.
\bibliographystyle{IEEEtran}
\bibliography{myreference}

\end{document}